\documentclass[aps,prb,twocolumn,superscriptaddress,floatfix,showpacs]{revtex4}

\usepackage{amsmath}
\usepackage{amsfonts}
\usepackage{graphicx}
\usepackage{bm}

\begin{document}

\title{Electronic Correlations in Double Quantum Dots}

\author{Jaime \surname{Zaratiegui Garc\'\i a}}
\affiliation{Department of Physical Sciences/Theoretical Physics,
  P.O. Box 3000, FIN-90014, University of Oulu, Finland}
\email[]{jaime.zaratiegui@oulu.fi}

\author{Pekka \surname{Pietil\" ainen}}
\affiliation{Department of Physical Sciences/Theoretical Physics,
  P.O. Box 3000, FIN-90014, University of Oulu, Finland}

\author{Hong-Yi \surname{Chen}}
\affiliation{Department of Physics and Astronomy, University of
  Manitoba, Winnipeg, Canada R3T 2N2}

\author{Tapash \surname{Chakraborty}}
\affiliation{Department of Physics and Astronomy, University of
  Manitoba, Winnipeg, Canada R3T 2N2}

\date{\today}

\begin{abstract}
We present a study of the electronic structure of two laterally coupled Gaussian quantum dots filled with two particles. The exact diagonalization method has been used in order to inspect the spatial correlations and examine the particular spin singlet-triplet configurations for different coupling degrees between quantum dots. The outcome of our research shows this structure to have highly modifiable properties promoting it as an interesting quantum device, showing the possible use of this states as a quantum bit gate.
\end{abstract}


\maketitle

The growing significance of mesoscopic quantum dots (QDs) in the development of microelectronic devices attracts great attention, partly due to their promising technological utilization, but also due to their novel quantum properties. These two-dimensional nanostructures offer a high degree of flexibility in their manufacturing process giving as a result a very interesting combination of a highly controllable structure showing quantum effects. Recent advances in the manipulation of electrically defined QDs open a new line of investigation in coupled quantum dots (CQDs) and a very interesting chance for the realization of a solid state implementation of the basic components for quantum computing. The interest on CQDs in the field of quantum computing was brought over since Loss and DiVincenzo proposal~\cite{Loss:1998} on the possibility of an implementation based on the spin states of coupled single-electron quantum dots. Recent studies and experiments~\cite{Hu:2000,Harju:2002,Krenner:2005,Ortner:2005,Helle:2005} have proven the feasibility of using the spin degree of freedom in vertically and laterally coupled few-electron QDs as systems in which interactions can be electro-optically tuned and therefore they can potentially taken as candidates for performing quantum operations.

It is the purpose of the present paper to present an accurate theoretical study focused on the electronic structure of the two-particle GDQD in the absence of a magnetic field. The exact diagonalization method~\cite{Davidson:1993} has been used in order to solve the interacting many-body system as it is a tool of proven reliability for few-particle problems~\cite{Chakraborty:1999}. Electron-electron correlation effects are studied for the different spin superposition states and for a variety of configurations, from a weakly interacting laterally coupled quantum dot molecule to a strongly overlapped situation. The choice of an anisotropic Gaussian confining potential makes it possible to describe a more realistic situation.~\cite{Hu:2000} We will show that not only in vertically coupled QDs interactions can be tuned but also in lateral QDs this effect can be achieved; furthermore, this systems could actually be used as dynamically tunable two-level systems.

As shown in our previous research~\cite{Chen:2007}, the special behavior of a GDQD filled with two electrons makes it a very promising arrangement for the creation of a dynamically tunable quantum gate. As noted by previous research~\cite{Hu:2000,Wensauer:2000,Harju:2002}, one of the most attractive setup among few-particle lateral quantum dots is the two-electron case in which the singlet and triplet states can be tuned to be bound into a fourfold degenerate ground state or, depending in the double QD parameters, artificially introducing an energy gap between the singlet and triplet states. This particular characteristic is very similar to that already observed in a two-minima QD subjected to a magnetic field~\cite{Helle:2005}. The possibility of using two coupled quantum dots containing two electrons as a charge qubit by dynamically tuning the interdot interaction has been recently proposed by Weiss \emph{et al.}~\cite{Weiss:2006}. Despite them using two vertically coupled QDs, our system is physically similar and it could also be employed in developing a charge qubit.

\begin{figure*}
\includegraphics[width=0.98\textwidth]{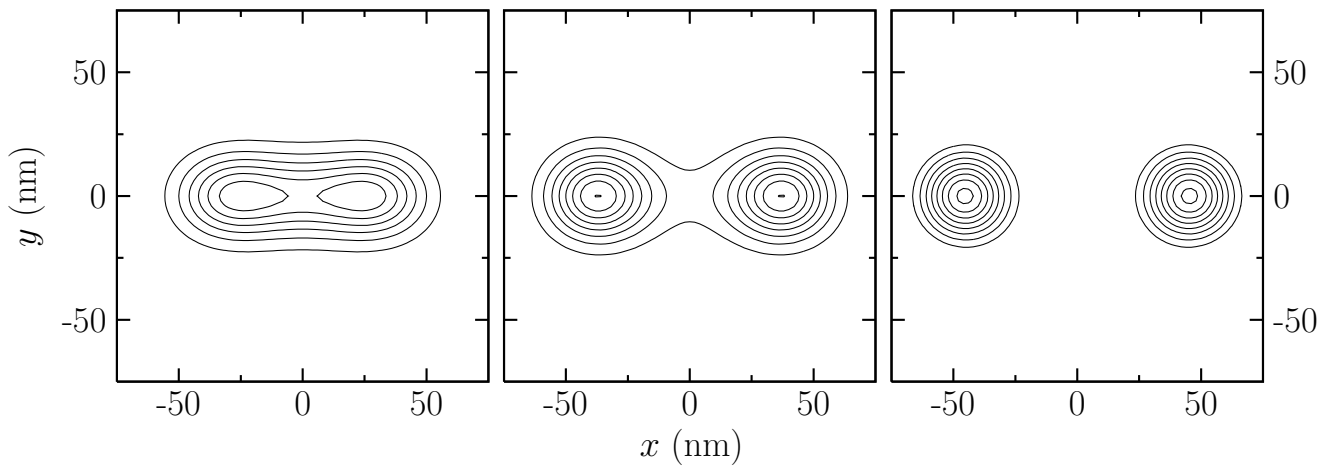}
\caption{Electron density contours for a GDQD with $V_b=4.0$ meV (left), 8.5 meV (center) and 14.0 meV (right). Left and center configurations show a clear coupling between both dots, whereas circular symmetry in each one of the dots is practically restored when $V_b$ is set to 14.0 meV. The point of maximum density is always placed at $y=0$ due to symmetry and, along the $x$-axis is placed at  $\pm 24$ nm, $\pm 37$ nm and $\pm 45$ nm from left to right. These points will be used when calculating conditional probabilities. We must note that setting $V_b$ to a sufficiently large and negative value, so that $V_0$ is negligible compared with $V_b$, we could restore a case in which only a single minima Gaussian QD is obtained.}
\label{fig:rho}
\end{figure*}

In our model, the single-particle two-dimensional Hamiltonian for a particle subjected to a magnetic field is given by
\begin{equation}
\mathcal{H}_0 = \sum_i^N
\frac{\left(\mathbf{p}_i-e\mathbf{A}_i/c\right)^2}{2m^\ast}
+ V_\mathbf{G}(\mathbf{r}_i) + \frac12 g^\ast\mu_\mathrm{B} B\sigma_z,
\end{equation}
where $m^\ast$ is the effective mass of the electron, $e=-|e|$ is the electron charge, $g^\ast$ is the effective Land\'e $g$-factor and $V_\mathrm{G}(\mathbf{r})$ is the external confining potential. We have used the model by Hu and Das Sarma~\cite{Hu:2000} in order to describe a realistic trapping potential. In this case, $V_\mathrm{G}$ is formed by three anisotropic Gaussians, two of them modeling the lateral quantum dots and the third one controlling the height of the potential barrier between them. More precisely, $V_\mathrm{G}(x, y)$ can be written as
\begin{eqnarray}
V_\mathrm{G} (x,y)& = & V_o\left[ \exp\left(-\frac{(x-x_0)^2}{l_x^2}\right)\exp\left(-\frac{(x+x_0)^2}{l_x^2}\right)\right] \nonumber \\&   & \times \exp\left(-\frac{y^2}{l_y^2}\right)+ V_b \exp\left(-\frac{x^2}{l_{bx}^2}-\frac{y^2}{l_{by}^2}\right),
\label{eq:VG}
\end{eqnarray}
with all parameters being positive except $V_0$, which controls the strength of the lateral QDs. $x_0$ regulates their relative distance. As we have mentioned before, we must note that the QDs studied in this paper do not exhibit circular symmetry as, in general, $l_x\neq l_y$ and $l_{bx}\neq l_{by}$. In this model, the parameter $V_b$ independently modulates the height of the interdot potential barrier, effectively controlling the interdot distance. It is important to note that solely varying the parameter $V_b$ we can smoothly jump from a single quantum dot filled with two electrons or, in the opposite case, with two independent quantum dots each of them occupied by a single electron. This is a very interesting parameter as it could be dynamically modified in an electrically defined QD giving the chance to exploit all of its properties. For the many-body Hamiltonian, it will be enough to include the Coulomb electrostatic potential, given by
\begin{equation}
V_\mathrm{C}(\mathbf{r}_1,\mathbf{r}_2)=\frac{e^2}{\epsilon'}r_{12}^{-1}.
\end{equation}

We have used the \emph{exact diagonalization} (ED) method in order to solve the Hamiltonian problem. Under this approach, we can write the eigenfunctions in the form of two-component spinors
\begin{equation}
\psi=\sum_{\kappa, s} c_{\kappa,s}\phi_{\kappa}\chi^s,
\end{equation}
where $\chi^\uparrow$ and $\chi^\downarrow$ are the eigenvectors of the $s_z$ operator and the index $\kappa$ points to a set of quantum numbers. The spatial wave functions chosen are the complete set of the solution of a two-dimensional anisotropic harmonic oscillator with characteristic lengths $a_x$ and $a_y$. The many-body interacting Hamiltonian is formed by including the two-body matrix elements of the Coulomb interaction, in the same fashion as in Ref.~\cite{Chen:2007}. Given this basis, the Hamiltonian matrix elements can be written in a closed and analytic form.~\cite{Chen:2007}

Using the second quantization formalism~\cite{Gross:1991} one can define the \emph{two-body density operator} or the \emph{pair correlation function} as
\begin{equation}
\rho(\mathbf{x}_1,\mathbf{x}_2)= \sum_{ijkl}\phi^\ast_{i}(\mathbf{x}_1)\phi^
\ast_{j}(\mathbf{x}_2)\phi_{k}(\mathbf{x}_2)\phi_{l}(\mathbf{x}_1)
\end{equation}where $\mathbf{x}=\{\mathbf{r},\zeta\}$ refers simultaneously to spatial and spin variables. If we are interested only in spatial correlations, we can sum over the spin degree of freedom to obtain
\begin{eqnarray}
\rho(\mathbf{r}_1,\mathbf{r}_2)&=&\rho_{\uparrow\uparrow}(\mathbf{r}_1,\mathbf{r}_2)+\rho_{\uparrow\downarrow}(\mathbf{r}_1,\mathbf{r}_2)\nonumber\\& & +\rho_{\downarrow\uparrow}(\mathbf{r}_1,\mathbf{r}_2)+\rho_{\downarrow\downarrow}(\mathbf{r}_1,\mathbf{r}_2),
\end{eqnarray}
where $\rho_{\uparrow\uparrow}(\mathbf{r}_1,\mathbf{r}_2)$ is the pair correlation function for for two spin up electrons and $\rho_{\uparrow\downarrow}(\mathbf{r}_1,\mathbf{r}_2)$ is the same quantity but for two anti-parallel particles.

\begin{figure}
\includegraphics[width=0.5\textwidth]{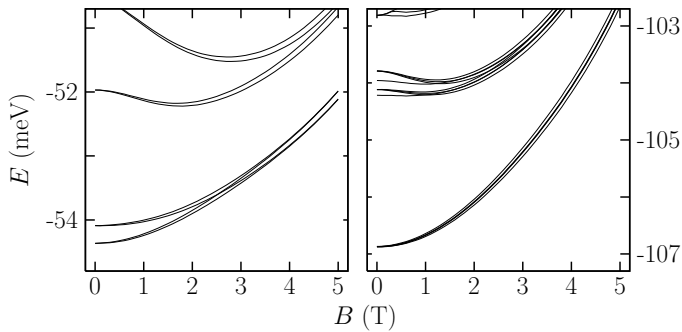}
\caption{Energy dependence with magnetic field for a single- (left) and two-
electron (right) GDQD. Interdot potential strength is fixed to $V_b=8.5$ meV. 
The two lowest energy branches in the case of a single-electron are formed by 
two levels each, merging as magnetic field increases.}
\label{fig:energy}
\end{figure}

It is our purpose to investigate the electronic structure of this system under different conditions. For accomplishing this objective, we have chosen the following set of shared parameters for the confining potential $V_\mathrm{G}$: $V_0=50$ meV, $x_0=54$ nm, $l_x=l_y=80$ nm, $l_{bx}=40$ nm and $l_{by}=50$ nm. Material parameters are those appropriate of GaAs, $m^\ast=0.067 m_\mathrm{e}$, $\epsilon'=13.1$, $g^\ast=-0.44$. Given all these values, we have done different calculations for the parameter $V_b$, from a strongly bound double QD in which $V_b=4.0$ meV to a case in which both QDs are nearly isolated from each other being $V_b=14.0$ meV, going trough an intermediate configuration in which $V_b=8.5$ meV. We will consider this intermediate setup as our base case. For this case, we can see in Fig.~\ref{fig:energy} the energy dependence on magnetic field for a single- and double-electron configuration. There, we can appreciate that in the single-particle case, ground state at $B=0$ T is a doubly degenerate level due to spin degree of freedom. This degeneracy is temporarily broken by magnetic field until reaching a sufficiently large $B$, which couples ground state and second degenerate state and makes them degenerate again. In the case of a two-electron GDQD, we can see that magnetic field splits level spacing monotonically. It is difficult to appreciate by naked eye, but energy difference between the first four states at $B=0$ T is of the order of the numerical precision used for this simulation. If we consider the same quantity being $V_b=14.0$ meV, this difference will be well below our numerical precision, but for the case $V_b=4.0$, the energy difference is not inappreciable anymore.

\begin{figure*}[!htp]
\includegraphics[width=0.89\textwidth]{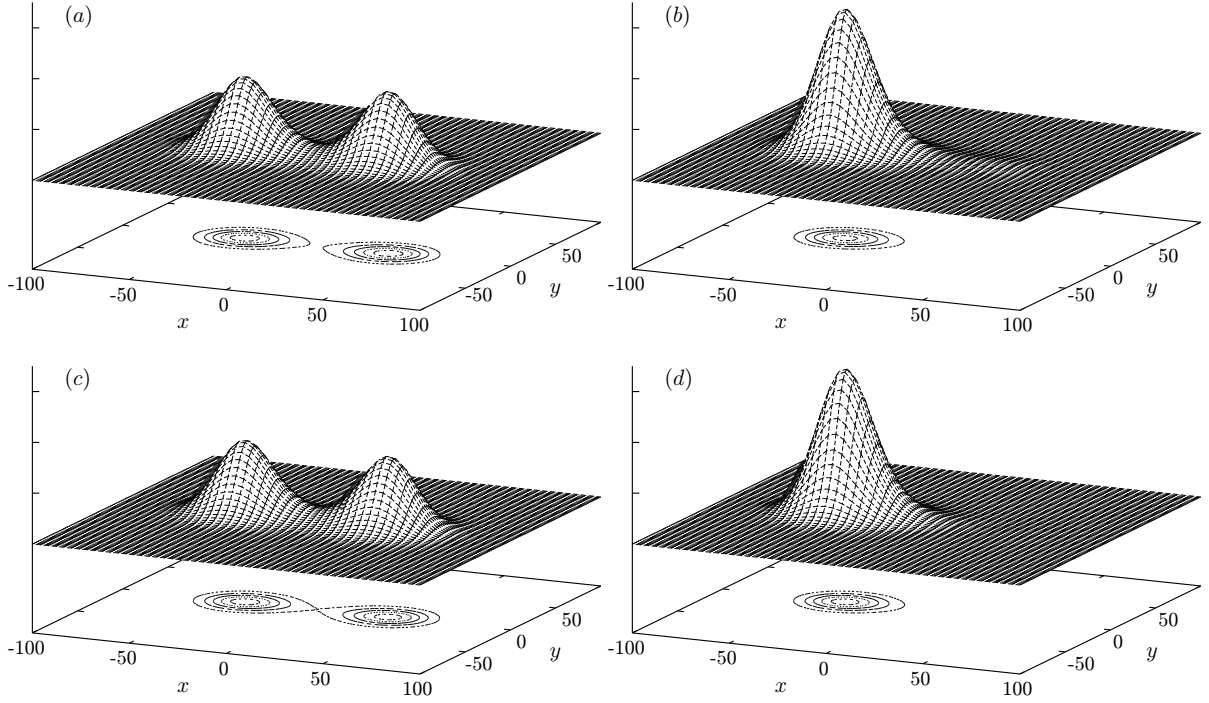}
\caption{Pair-correlation function $\rho(\mathbf{r},\mathbf{r}')$ for the GDQD 
with $V_b=8.5$ meV and $\mathbf{r}'=(x', y')=(37, 0)$ nm. Figure $(a)$ shows 
ground state conditional density and figures $(b)$, $(c)$ and $(d)$ correspond 
to first, second and third excited states. Scale in $x$ and $y$ axes is measured in nm.}
\label{fig:corr8_5}
\end{figure*}

\begin{figure*}[!hbtp]
\includegraphics[width=0.89\textwidth]{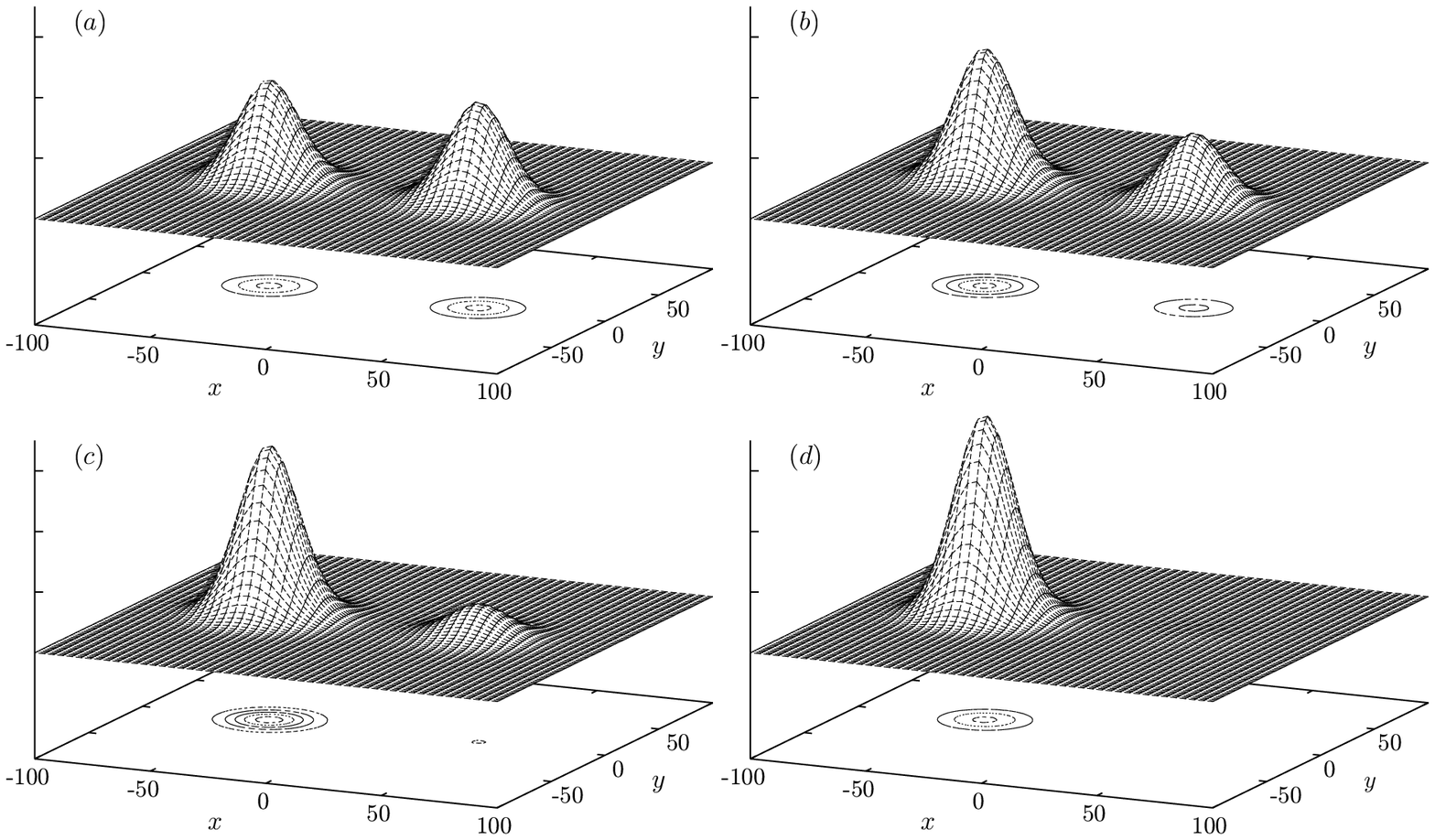}
\caption{Pair-correlation function $\rho(\mathbf{r},\mathbf{r}')$ for the GDQD 
with $V_b=14.0$ meV and $\mathbf{r}'=(x', y')=(45, 0)$ nm. Figure $(a)$ shows 
ground state conditional density and figures $(b)$, $(c)$ and $(d)$ correspond 
to first, second and third excited states. Scale in $x$ and $y$ axes is measured in nm.}
\label{fig:corr14_0}
\end{figure*}

The total electron density $\rho(\mathbf{r})=\rho_\downarrow(\mathbf{r})+\rho_\uparrow(\mathbf{r})$ for the ground state is plotted in Fig.~\ref{fig:rho}. We can appreciate a clear separation of the quantum dots as $V_b$ is increased, in fact, in the case for higher $V_b$, we almost recover two clearly differentiated and circularly symmetric QDs. Interestingly, the effect in electron density when modifying the parameter $V_b$ is similar to that observed by a variable magnetic field noted by Marlo and Helle~\cite{Marlo:2003,Helle:2005}. From these figures we can find out the maximum electron density points in order to use them when calculating the correlations between electrons. Due to axial symmetry, these point will be placed at $y=0$ nm, and they will be symmetric with respect to $x$-axis. From left to right, the abscissa is 24, 37 and 45 nm. Given this values, we have plotted Figs.~\ref{fig:corr8_5} and~\ref{fig:corr14_0}.

Results for electron-electron correlations when $V_b=8.5$ meV are plotted in Fig.~\ref{fig:corr8_5}. One can appreciate a clear symmetric-antisymmetric pattern in the distribution of the energy levels, this can be explained by the spin change of parity in the spinors. For example, in Fig.~\ref{fig:corr8_5}~(a), we can find the conditional probability for the ground state, it shows a spatially symmetric distribution, this is due to the equal contribution of $\rho_{\uparrow\downarrow}$ and $\rho_{\downarrow\uparrow}$ terms into the total correlation $\rho(\mathbf{r}_1, \mathbf{r}_2)$ while the numerical contribution of the remaining terms is numerically negligible. This symmetry is also observed in subfigure \ref{fig:corr8_5}~(c), corresponding to the second excited state. In opposition, Figs. \ref{fig:corr8_5}~(b) and \ref{fig:corr8_5}~(d), those corresponding to first and third excited states, are fundamentally composed by the spin-symmetrized $\rho_{\uparrow\uparrow}$ and $\rho_{\downarrow\downarrow}$ terms. As we are dealing with an even ground state has a spatially symmetric distribution as well as second excited state,

The energy spread between the first four lowest lying levels, \emph{i.e.} singlet and triplet states, varies dramatically as a result of modifying the interdot barrier height.~\cite{Chen:2007} Beginning with the case $V_b=4.0$ meV in which $\Delta E=65.30$~$\mu$eV and ending at $V_b=14.0$ meV we find that $\Delta E=0.12$ $\mu$eV. In the intermediate step, $V_b=8.5$ meV the energy spread is 5 $\mu$eV. The numerical error derived from our diagonalization routines for the highest barrier case is estimated to be of the same magnitude as the energy difference between singlet and triplet states, this impedes us to clearly discern between the four energy levels and indicates us that we must take them as a group of degenerate levels. This situation contrasts with the cases with a lower interdot barrier, in which energy separation is clear and there is no evidence of energy level degeneracies to appear.

In summary, we have studied the effect of electrostatic interactions in a two-electron laterally coupled double quantum dot. The results of our exact diagonalization calculations show a peculiar singlet-triplet level merging as coupling strength is screened by a interdot barrier. Further analysis of the electronic wave functions such as calculating conditional probabilities of the lowest energy levels, show the change produced in the spatial and spin degrees of freedom.

The work has been supported in part by the Canada Research Chair program and a NSERC Discovery grant. We thank the authors of Ref.~\onlinecite{Weiss:2006} for bringing that paper to our attention.



\bibliography{dblcorr}

\end{document}